\documentclass[useAMS,usenatbib,a4]{mn2e}

\usepackage[utf8]{inputenc}
\usepackage{amsmath}
\usepackage{amssymb}
\usepackage{graphicx}
\usepackage{hyperref}
\usepackage{times}

\newcommand{\dd}{\,\mathrm{d}\,}
\newcommand{\pp}{\partial}

\newcommand{\D}{\displaystyle}

\renewcommand{\vec}[1]{\bmath{#1}}
\newcommand{\bhat}[1]{\hat{\bmath{#1}}}

\title{Revisiting the Flowers-Ruderman instability of magnetic stars}
\author[Pablo Marchant, Andreas Reisenegger and Taner Akg\"un]{Pablo Marchant$^{\text{1}}$, Andreas Reisenegger$^{\text{1}}$ and Taner Akg\"un$^{\text{1,2}}$\\
$^{\text{1}}$Departamento de Astronom\'ia y Astrof\'isica, Pontificia Universidad Cat\'olica de Chile, Casilla 306, Santiago, Chile\\
$^{\text{2}}$Barcelona Supercomputing Center - Centro Nacional de Supercomputaci\'{o}n, C/ Gran Capit\`{a} 2-4, Barcelona, 08034, Spain\\
E-mails: pmarchan@astro.puc.cl (PM); areisene@astro.puc.cl (AR); akgun@astro.cornell.edu (TA)
}
\begin{document}
\maketitle
\begin{abstract}
In 1977, Flowers and Ruderman described a perturbation that destabilises a purely dipolar magnetic
field in a fluid star. They considered the effect of cutting the star in half along a plane containing the
symmetry axis and rotating each half by $90\degr$ in opposite directions, which would cause the energy
of the magnetic field in the exterior of the star to be greatly reduced, just as it happens with a pair
of aligned magnets. We formally solve for the energy of the external magnetic field and check that it
decreases monotonically along the entire rotation. We also describe the instability using perturbation
theory, and show that it happens due to the work done by the interaction of the magnetic field with
surface currents. Finally, we consider the stabilising effect of adding a toroidal field by studying the
potential energy perturbation when the rotation is not done along a sharp cut, but with a continuous
displacement field that switches the direction of rotation across a region of small but finite width. Using
these results, we estimate the relative strengths of the toroidal and poloidal fields needed to make the
star stable to this displacement and show that the energy of the toroidal field required for stabilisation is much
smaller than the energy of the poloidal field. We also show that, contrary to a common argument, the Flowers-Ruderman instability cannot be applied many times in a row to reduce the external magnetic energy indefinitely.
\end{abstract}
\section{Introduction}
Large-scale magnetic fields are known to be present in a wide variety of stellar objects, meaning that in these stars the dipole component (together perhaps with some other low-order multipoles) is not much weaker than the rms surface field. The initial discovery of such fields was in Ap stars (\citealt{bab+47}). Since then, they have been observed or inferred to exist in white dwarfs, neutron stars, upper-main-sequence stars, and in the central stars of planetary nebulae. These fields appear to be long-lived, since they do not evolve on a timescale accessible to observations (see, for instance, \citealt{bra+10}).

A common feature of all these objects is that, over most of their interior, they are stably stratified. White dwarfs and neutron stars have no significant convective regions\footnote{Recently formed neutron stars are only convective for some seconds, and white dwarfs have a thin convective region near their surface.}, while upper-main-sequence stars only have a small convective core. Dynamo effects are therefore expected to be irrelevant in keeping the strength of the magnetic field constant. The maximum surface magnetic fluxes observed in all these objects are similar, $\Phi_{\mathrm{max}}=\pi R^2 B_{\mathrm{max}}\sim 10^{27.5}\;\mathrm{G\;cm^2}$, where $B_{\mathrm{max}}$ is the highest surface dipole strength detected in each class of objects. These two features are considered compelling arguments in favour of flux freezing during stellar evolution, although magnetic flux is certain to be lost in supernova explosions, so it is plausible that the magnetic field is regenerated somehow in a young neutron star (\citealt{thodun+93}).

The ratio of fluid to magnetic pressure is (\citealt{rei+09})
 \begin{eqnarray}
 \beta=\frac{8\pi P}{B^2}\sim \frac{8\pi^3 G M^2}{\Phi^2}\sim 3\times 10^6 \left(\frac{M}{M_{\odot}}\right)^2 \left(\frac{\Phi}{\Phi_{\mathrm{max}}}\right)^{-2},  
 \end{eqnarray}
 which is a very large number even for the most strongly magnetised stars, unless the internal magnetic fields are substantially larger than the surface magnetic fields. Since this ratio is so high, we do not expect these fields to significantly modify the hydrostatic structure of the stars (i.e. their internal density and temperature profiles). However, they can play a major role in their evolution, e.g., through the transport of angular momentum (see, for instance, \citealt{hewosp+00}).

Even though these long-lived fields have been known to exist for more than half a century, it has not been possible to find an analytic model for a field that is in a stable equilibrium. However, stable configurations that do not seem to evolve on timescales comparable to the Alfvén crossing time have been found to exist via numerical calculations (\citealt{braspru+04}), where an initially random field usually evolves into an approximately axisymmetric configuration that is a combination of toroidal and poloidal components of similar energies (for axisymmetric fields, toroidal and poloidal
refer to the azimuthal and meridional components of the field, respectively).

The stability of purely poloidal or purely toroidal fields has also been studied in the past. \citet{tay+73}, using the energy method, suggested that every purely toroidal field is unstable on an Alfv\'en timescale, independent of the strength of the field. \citet{martay+73,martay+74} and independently \citet{wri+73} discovered that purely poloidal fields with closed lines contained inside the star are affected by an instability that is very similar to the kink instabilities in a Z-pinch.

\begin{figure}
\centering
\includegraphics[height=2in]{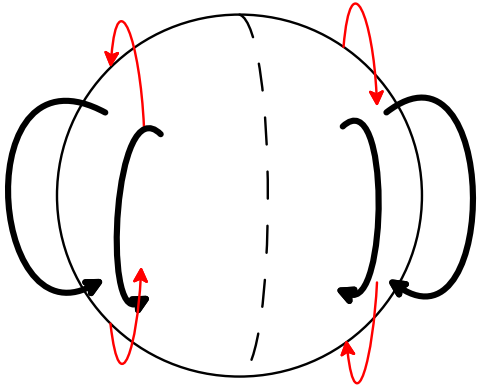}
\caption{Displacement produced by the Flowers-Ruderman instability. The thick arrows are field lines from the dipole, the dashed line shows the plane along which the star is cut, and both halves are rotated in opposite directions as shown by the thin arrows.}
\label{fandrfig}
\end{figure}

A simple argument given by \citet{florud+77} shows that any purely poloidal field with field lines extending outside the star should be unstable. If the initial configuration is such that the external field resembles a dipole, cutting the star in half and rotating each half by $90\degr$ in opposite directions (as shown in Fig. \ref{fandrfig}) would greatly reduce the dipole component of the field, leading to a magnetic field with less energy. However, neither Flowers and Ruderman nor anyone else have given a formal proof of this argument.

In the numerical simulations of \citet{bra+09}, instabilities related to the poloidal and toroidal components of the field are studied. Using the stable configurations found after simulating the evolution of random fields, Braithwaite used different ratios of poloidal to total energy of the magnetic field, $E_{\mathrm{P}}/E$, and found the field to be stable for $0.056<E_{\mathrm{P}}/E<0.8$. The field became unstable for $E_{\mathrm{P}}/E > 0.8$, with an $m=2$ mode that seems to consist mostly of displacements in latitude of the fluid. For $E_{\mathrm{P}}/E>0.9$, modes with higher $m$ became unstable, as would be expected since these modes have to overcome a higher resistance from the toroidal field. These modes resemble kink instabilities, as was mentioned before.

The structure of this paper is the following: In \S\ref{far::proofEx}, we formally prove the Flowers-Ruderman instability for a pure dipole field. To do so, we explicitly calculate the energy of the external magnetic field as a function of the angle of rotation of each half of the star, and see that it is a monotonically decreasing function. In \S\ref{far::pertu} we treat the problem using perturbation theory for a particular family of fields. In \S\ref{csfar}, we estimate the stabilising effect of a toroidal field when the perturbation is not done with a sharp cut through the star, but rather with a displacement field that switches continuously from one direction of rotation to the other, over a thin but finite region. In \S\ref{far::succesive}, we show that, contrary to what has been claimed previously in the literature, higher order multipoles cannot be achieved with successive cuts in different directions, and in \S\ref{conclusions}, we present the conclusions of our work.

\section{Proof of the Flowers-Ruderman instability by an exact evaluation of the energy}\label{far::proofEx}
We consider the star to be non-rotating and completely surrounded by vacuum. Inside the star, the magnetic field is axisymmetric and purely poloidal, and outside the star the field is a pure dipole. If we completely ignore the effects of the magnetic field on the hydrostatic structure of the star (an assumption we justify in \S\ref{far::sphere}), then the star should be perfectly spherical, and when the Flowers-Ruderman instability takes place, each half of the star rotates as a rigid solid. Since in stellar interiors the magnetic Reynolds number is much larger than $1$, field lines will be dragged by the fluid without modifying the magnitude of the magnetic field at each point, and thus the internal magnetic energy of the star will not be modified in the process\footnote{It is important to note that the plane that cuts the star cannot cross any field lines, otherwise, to rotate each half these field lines should be cut, and that is not possible.}. Therefore, we are only interested in the energy of the external magnetic field, and we now proceed to prove that this energy is in fact monotonically reduced by performing the displacement suggested by Flowers and Ruderman.
\subsection{Exterior energy of an arbitrary magnetic field}\label{far::extE}
To start, we must obtain the magnetic field outside the star, given the field on its surface. Because outside the star there are no currents, we have $\nabla\times\vec{B}=0$ and therefore $\vec{B}=\nabla\Psi$. Since $\nabla\cdot\vec{B}=0$, $\Psi$ must satisfy Laplace's equation,
\begin{eqnarray}
    \nabla^2\Psi=0.
\end{eqnarray}
The solution to this equation in spherical coordinates that is of physical significance to us is
\begin{eqnarray}
\Psi(r,\theta,\phi)=\sum_{l=1}^{
\infty}\sum_{m=-l}^{l}\frac{a_{lm}}{r^{l+1}}Y_{lm}(\theta,\phi),\label{Psi}
\end{eqnarray}
where the term with $l=0$ is excluded since it corresponds to a magnetic monopole. The $a_{lm}$ are obtained by requiring the field to have a continuous normal component at the surface,
\begin{eqnarray}
a_{lm}=-\frac{R^{l+2}}{l+1}\int_{4\pi}Y_{lm}^*(\theta,\phi)(B_r)_{r=R}\dd\Omega,\label{alm}
\end{eqnarray}
where $B_r=\vec{B}\cdot\bhat{r}$ is the radial component of the field. Now, the magnetic energy inside the star should not change, since the field only rotates while keeping its magnitude. However, the exterior field changes significantly. Thus, the variation of the magnetic energy can be obtained just by computing the variation outside of the star. The exterior magnetic energy is obtained from
\begin{eqnarray}
 \D E=&\D\hspace{-0.1in}\int_V \frac{B^2}{8\pi} \dd V=\int_V  \frac{(\nabla\Psi)^2}{8\pi} \dd V=\D\int_V \frac{\nabla\cdot(\Psi\nabla\Psi)}{8\pi} \dd V
\end{eqnarray}
where $V$ covers all space outside the star, and we used the fact that $\Psi$ satisfies Laplace's equation. Using the divergence theorem, the energy can be expressed as a surface integral, with a normal pointing into the star:\footnote{The term corresponding to the surface at infinity vanishes. This is because the term in $\Psi$ that decreases most slowly with $r$ goes like $r^{-2}$, so $(\Psi\nabla\Psi)\cdot\dd \vec{S}$ goes like $r^{-3}$ and vanishes in the limit $r\rightarrow\infty$.}
\begin{eqnarray}
 E=\frac{1}{8\pi}\oint_S  (\Psi\nabla\Psi)_{r=R}\cdot \dd \vec{S}.
\end{eqnarray}
Since we consider the star to be perfectly spherical,\linebreak $\nabla\Psi\cdot\dd \vec{S}=-R^2(B_r)_{r=R}\dd\Omega$, and consequently
\begin{eqnarray}
 E=-\frac{R^2}{8\pi}\int_{4\pi}(\Psi B_r)_{r=R}\dd\Omega=-\frac{R^2}{8\pi}\int_{4\pi}(\Psi^*B_r)_{r=R}\dd\Omega
\end{eqnarray}
where in the last step, we used the fact that $\Psi$ is real and set it equal to its conjugate. Replacing the expression for $\Psi$, as given by eq. (\ref{Psi}), we get the following result:
\begin{eqnarray}
\begin{aligned}
 \D E=&\D-\frac{R^2}{8\pi}\sum_{lm}\frac{a_{lm}^*}{R^{l+1}}\int_{4\pi}Y_{lm}^*(\theta,\phi)(B_r)_{r=R}\dd\Omega\\
=&\D\frac{R^3}{8\pi}\sum_{lm}\frac{1}{(l+1)}\left|\frac{a_{lm}(l+1)}{R^{l+2}}\right|^2,
\end{aligned}
\end{eqnarray}
where we used eq. (\ref{alm}) to express the integral in terms of the $a_{lm}$. If we define $c_{lm}$ as
\begin{eqnarray}
 c_{lm}=-\frac{a_{lm}(l+1)}{R^{l+2}}=\int_{4\pi}Y_{lm}^*(\theta,\phi)(B_r)_{r=R}\dd\Omega,\label{clmdef}
\end{eqnarray}
the energy of the external magnetic field is
\begin{eqnarray}
 E=\frac{R^3}{8\pi}\sum_{lm}\frac{|c_{lm}|^2}{l+1}.\label{EFE}
\end{eqnarray}
Another useful result is the radial component of the field at the surface expressed in terms of the $c_{lm}$ coefficients. This can be obtained from the radial component of $\nabla\Psi$, or by inverting eq. (\ref{clmdef}), and it is equal to
\begin{eqnarray}
 (B_r)_{r=R}=\sum_{lm}c_{lm}Y_{lm}.\label{radialb}
\end{eqnarray}

\subsection{Proof that the final energy is less than the initial one}\label{far::less}
The results contained in eq. (\ref{EFE}) can be used to prove that the energy is effectively reduced when one half of the star is rotated with respect to the other. To do so, let us define a quantity $\Upsilon$ as
\begin{eqnarray}
\Upsilon=\frac{R^3}{8\pi}\int_{4\pi}(B_r)_{r=R}^2 \dd\Omega\label{Upsilon}.
\end{eqnarray}
This quantity will be conserved when the star is cut in half and rotated. So, using the superscripts $i$ and $f$ to denote initial and final states, $\Upsilon^\mathrm{i}=\Upsilon^\mathrm{f}$. If we use the result of eq. (\ref{radialb}) to express the terms of $(B_r)_{r=R}^2=(B_rB_r^*)_{r=R}$, we get
\begin{eqnarray}
\Upsilon&=&\D\frac{R^3}{8\pi}\sum_{lm}\sum_{l'm'}c_{lm}c_{l'm'}^*\int_{4\pi}Y_{lm}Y_{l'm'}^*\dd\Omega\\
&=&\D\frac{R^3}{8\pi}\sum_{lm}|c_{lm}|^2.
\end{eqnarray}
By rewriting $\Upsilon^\mathrm{i}=\Upsilon^\mathrm{f}$, we obtain
\begin{eqnarray}
\sum_{lm}|c_{lm}^\mathrm{f}|^2=\sum_{lm}|c_{lm}^\mathrm{i}|^2.\label{upsilonconv}
\end{eqnarray}
If the initial external field is a dipole field, the only nonzero $c_{lm}^\mathrm{i}$ is $c_{10}^\mathrm{i}$. Considering this,
\begin{eqnarray}
\sum_{lm} |c_{lm}^\mathrm{f}|^2=|c_{10}^\mathrm{i}|^2.
\end{eqnarray}
From here, using eq. (\ref{EFE}) we get that
\begin{eqnarray}
E_\mathrm{f}\le \frac{R^3}{8\pi}\sum_{lm}\frac{|c_{lm}^\mathrm{f}|^2}{2}=\frac{R^3}{8\pi}\frac{|c_{10}^\mathrm{i}|^2}{2}=E_\mathrm{i}.\label{ineki}
\end{eqnarray}
Thus, the final state will have less or equal energy than the initial one. The equality would hold if and only if the $c_{lm}^\mathrm{f}$ are equal to zero when $l\neq 1$, which is not the case since the severe discontinuity that is produced cannot be resolved into an expansion of spherical harmonics with a finite number of terms.

It is important to note, however, that we only proved that the magnetic energy of any final state after cutting the star and rotating it is less than the initial energy of the dipole field. We have yet to prove that the energy is monotonically decreasing for the entire rotation. So, up to this point, we could expect the minimum energy to be present at some intermediate point in the rotation, and not after the rotation has been completed.

\subsection{Proof that the energy decreases monotonically}\label{far::mon}
\begin{figure}
\centering
\includegraphics[height=2.5in]{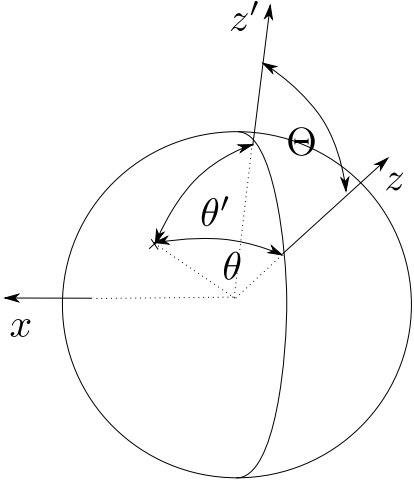}
\caption{Relationship between the coordinates $\theta$ and $\theta'$ used in \S\ref{far::mon}. Initially we have a spherical coordinate system in which the polar angle $\theta$ of a point in space corresponds to the angle formed with respect to the $z$ axis. A new axis $z'$ is defined by performing a right handed rotation by an angle $\Theta$ with respect to the $z$ axis, and the polar angle $\theta'$ of a point is defined as the angle formed with respect to the $z'$ axis.}
\label{rotyey}
\end{figure}

The initial field outside of the star is that of a dipole, so
\begin{eqnarray}
 (B_r)_{r=R}=B_0 \cos\theta,\label{brdipole}
\end{eqnarray}
where $B_0$ is the strength of the field at the poles. This can be written in the form of eq. (\ref{radialb}) as
\begin{eqnarray}
 (B_r)_{r=R}=B_0\sqrt{\frac{4\pi}{3}}Y_{10},
\end{eqnarray}
so from eq. (\ref{clmdef}), the only nonzero $c_{lm}$ is $c_{10}=B_0\sqrt{4\pi/3}$, and using eq. (\ref{EFE}), the external magnetic energy of this field can be evaluated as
\begin{eqnarray}
 E_\mathrm{e}=\frac{B_0^2R^3}{12}.\label{E0def}
\end{eqnarray}
Now, we rotate each half of the star by an angle $\Theta$ in opposite directions, 
\begin{eqnarray}
 (B_r)_{r=R}=
\begin{cases}
B_0\cos[\theta'(\theta,\phi,\Theta)]& x>0\\
B_0\cos[\theta'(\theta,\phi,-\Theta)]& x<0\\
\end{cases}
\end{eqnarray}
where $\theta'$ corresponds to the polar angle in a spherical coordinate system that rotates together with each half of the star, as illustrated in Fig. \ref{rotyey}.
It is fairly straightforward to show that
\begin{eqnarray}
 \cos[\theta'(\theta,\phi,\Theta)]=\cos\theta\cos\Theta-\sin\theta\sin\phi\sin\Theta,
\end{eqnarray}
so that the radial component of the field at the surface is
\begin{eqnarray}
 (B_r)_{r=R}=
\begin{cases}
B_0(\cos\theta\cos\Theta-\sin\theta\sin\phi\sin\Theta)& x>0\\
B_0(\cos\theta\cos\Theta+\sin\theta\sin\phi\sin\Theta)& x<0\\
\end{cases}
\end{eqnarray}
For this field, the $c_{lm}$ as given by eq. (\ref{clmdef}) are
\begin{eqnarray}
\begin{aligned}
 c_{lm}=&\cos\Theta B_0\int_{4\pi}\cos\theta Y_{lm}^*\dd \Omega\\
&+B_0\sin\Theta\left(\int_0^\pi\int_{\pi/2}^{3\pi/2}\right. \sin^2\theta\sin\phi  Y_{lm}^*\dd\phi\dd\theta \\
&\qquad\qquad\quad\;\;\left.-\int_0^\pi\int_{-\pi/2}^{\pi/2}\sin^2\theta\sin\phi Y_{lm}^*\dd\phi\dd\theta\right),
\end{aligned}
\end{eqnarray}
 and due to the symmetries of $Y_{lm}$, it can be shown that
\begin{eqnarray}
c_{lm}=
\begin{cases}
\D 2\sqrt{\pi/3} B_0\cos\Theta & l=1,m=0\\
\D 2w_{lm}B_0\sin\Theta & l\;\mathrm{and}\;m\;\mathrm{even}\\
\D 0 & \mathrm{otherwise}
\end{cases}\label{ceees}
\end{eqnarray}
where
\begin{eqnarray}
 w_{lm}=\int_0^\pi\dd\theta\int_{\pi/2}^{3\pi/2}\dd\phi\sin^2\theta\sin\phi Y_{lm}^*.
\end{eqnarray}
Using these values for the $c_{lm}$ combined with eq. (\ref{EFE}), the external magnetic energy is obtained as a function of $\Theta$,
\begin{eqnarray}
\begin{aligned}
 E=\cos^2\Theta E_\mathrm{e}+\sin^2\Theta\frac{R^3 B_0^2}{2\pi}\sum_{\substack{lm\\\mathrm{even}}}\frac{|w_{lm}|^2}{l+1}\\
=E_\mathrm{e}\left(\cos^2\Theta+\frac{6\sin^2\Theta}{\pi}\sum_{\substack{lm\\\mathrm{even}}}\frac{|w_{lm}|^2}{l+1}\right).
\end{aligned}
\end{eqnarray}
Defining
\begin{eqnarray}
 A=\frac{6}{\pi}\sum_{\substack{lm\\\mathrm{even}}}\frac{|w_{lm}|^2}{l+1}\label{Adef},
\end{eqnarray}
the energy can be rewritten in a compact form as
\begin{eqnarray}
 E=E_\mathrm{e}\left[1+\sin^2\Theta(A-1)\right]\label{energybz}.
\end{eqnarray}
Since the complete rotation is obtained with $\Theta=\pi/2$, the energy is a monotonic function of $\Theta$. If $A>1$, the energy will be an increasing function, but if $A<1$, it will be a decreasing function. However, from the results of \S\ref{far::less}, we already know that the final energy is smaller than the initial one, thus $A<1$ and the energy decreases monotonically along the entire rotation. Therefore, the Flowers-Ruderman instability is present in the case of the purely dipolar field.

Even though we already proved the existence of the instability, an estimate of $A$ is called for. To obtain this estimate, we consider the quantity
\begin{eqnarray}
 A_l=\frac{6}{\pi}\sum_{\substack{l'=2\\l'\;\mathrm{even}}}^l\sum_{\substack{m=-l'\\m\;\mathrm{even}}}^{l'}\frac{|w_{l'm}|^2}{l'+1}\label{gedeele},
\end{eqnarray}
which tends asymptotically to $A$ as $l$ increases. Also, due to the conservation of $\Upsilon$ described in \S\ref{far::less}, and the initial and final values of the $c_{lm}$, which can be obtained by setting $\Theta$ equal to zero or $90\degr$ respectively in eq. (\ref{ceees}), it can be seen that the $w_{lm}$ must satisfy
\begin{eqnarray}
\sum_{lm}|w_{lm}|^2=\frac{\pi}{3}.
\end{eqnarray}
Using this, we can obtain lower and upper bounds on the value of $A$,
\begin{eqnarray}
 A_l<A<A_l+\frac{6}{\pi(l+3)}\cdot\D\left(\frac{\pi}{3}-\sum_{\substack{l'=0\\l'\;\mathrm{even}}}^l\sum_{\substack{m=-l'\\m\;\mathrm{even}}}^{l'}|w_{l'm}|^2\right),
\end{eqnarray}
and since both limits tend asymptotically to $A$ as $l$ increases, this in principle can be used to evaluate $A$ to an arbitrary precision. By using $l=100$, one finds that
\begin{eqnarray}
 0.5463<A<0.5466.\label{Alimits}
\end{eqnarray}
The final energy of the system is $E_\mathrm{f}=AE_\mathrm{e}$, so the process nearly halves the energy of the external magnetic field. \citet{rob+81} gave a value of $A=0.577$, which is slightly higher than ours, but he did not describe exactly how he computed that number.
\subsection{Including higher order multipoles in the initial configuration}
Extending the previous analysis to consider the superposition of multipole components with the dipole component turns out to be a much more complicated problem. However, it is possible to get a simple answer if we only consider axisymmetric fields, and if we only care about the energy difference between the final and the initial state.

Consider an axisymmetric field that consists of a dipole and some higher order multipole, of fixed order $\hat{l}>1$,
\begin{eqnarray}
\begin{aligned}
  \Psi =& -\frac{c_{10}^\mathrm{i} R^3}{2r^2} Y_{10}-\frac{c_{\hat{l} 0}^\mathrm{i} R^{\hat{l}+2}}{\left(\hat{l}+1\right)r^{\hat{l}+1}} Y_{\hat{l}0},\\
\vec{B}=&\nabla \Psi,
\end{aligned}\label{axifield} 
\end{eqnarray}
where the superscript $i$ is used to denote that these coefficients represent the initial state. After the rotation, the field might not be axisymmetric, and it will be defined by some set of coefficients $c_{lm}^\mathrm{f}$. From eq. (\ref{EFE}), the final energy of the system is
\begin{eqnarray}
E_\mathrm{f}=\frac{R^3}{8\pi}\sum_{lm}\frac{|c_{lm}^\mathrm{f}|^2}{l+1}.\label{condiyey}
\end{eqnarray}

It can be shown that the final configuration has no dipole component, by proving that each multipole by itself does not produce a dipole component after the rotation. To do this, we consider one half of the star to be rotated by $180\degr$ instead of both halves being turned by $90\degr$ in opposite directions as was done in the previous section. The choice of how we perform the rotation is irrelevant to our calculation, since the energy associated with each multipole cannot depend on the particular spherical coordinate system chosen to perform the spherical harmonic expansion. However, the true physical process requires both halves to rotate in opposite directions due to the conservation of angular momentum.

In particular, it is evident that the final configuration has no dipole component when $\hat{l}$ is even, since in these cases $Y_{\hat{l}0}$ is symmetric with respect to the equator, and a rotation by $180\degr$ of one half of the star will leave the field just as it was at the beginning. For the case of odd $\hat{l}$, the dipole components $c_{1m}^\mathrm{f}$ that would be produced by that particular multipole after the rotation can be expressed from eq. (\ref{clmdef}) as
\begin{eqnarray}
  c_{1m}^\mathrm{f}= A_{\hat{l} m}\int_0^\pi P_{1m}(\cos\theta)\cdot P_{\hat{l}0}(\cos\theta)\sin\theta\dd\theta,
\end{eqnarray}
where the $P_{lm}$ are associated Legendre polynomials (\citealt{arf+05}) and $A_{\hat{l} m}$ contains the result of integrating over $\phi$, the normalising factors of the spherical harmonics, and an additional factor that relates to the strength of the multipole. Due to the orthogonality condition for the associated Legendre polynomials, $c_{10}^\mathrm{f}$ is equal to zero, and since $P_{1\pm1}(\cos\theta)\sim \sin\theta$ (except for a numerical constant), we have
\begin{eqnarray}
\begin{aligned}
  c_{1\pm1}^\mathrm{f}=&\D A_{\hat{l}\pm 1}\int_0^\pi P_{ \hat{l}0}(\cos\theta)\sin^2\theta\dd\theta\\
  =&\D A_{\hat{l}\pm 1}\int_{-1}^{1} P_{\hat{l}0}(x)\sqrt{1-x^2}\dd x=0,
\end{aligned}
\end{eqnarray}
where the equality to zero is due to the fact that the $P_{l0}(x)$ are odd functions of $x$ for odd $l$. Thus, the dipole component is zero after the rotation, so eq. (\ref{condiyey}) can be written as
\begin{eqnarray}
  E_\mathrm{f}=\frac{R^3}{8\pi}\sum_{l\ge2,m}\frac{|c_{lm}^\mathrm{f}|^2}{l+1}<\frac{R^3}{8\pi}\sum_{l\ge2,m}\frac{|c_{lm}^\mathrm{f}|^2}{3}.\label{finalE}
\end{eqnarray}
Due to the conservation of $\Upsilon$ shown in eq. (\ref{upsilonconv}),
\begin{eqnarray}
  \frac{R^3}{8\pi}\sum_{l\ge2,m}\frac{|c_{lm}^\mathrm{f}|^2}{3}&=&\D\frac{R^3}{8\pi}\left(\frac{|c_{10}^\mathrm{i}|^2}{3}+\frac{|c_{\hat{l}0}^\mathrm{i}|^2}{3}\right)\\
  &=&\D\frac{2}{3}E_\mathrm{D}+\frac{\hat{l}+1}{3}E_\mathrm{M},\label{superfinalE}
\end{eqnarray}
where $E_\mathrm{D}$ and $E_\mathrm{M}$ correspond to the initial energies of the dipole and the multipole components,
\begin{eqnarray}
  E_\mathrm{D}=\frac{R^3}{8\pi}\frac{|c_{10}^{i}|^2}{2},\quad E_\mathrm{M}=\frac{R^3}{8\pi}\frac{|c_{\hat{l} 0}^{i}|^2}{\hat{l}+1},
\end{eqnarray}
and the initial energy of the system is simply $E_\mathrm{i}=E_\mathrm{D}+E_\mathrm{M}$. Combining eqs. (\ref{finalE}) and (\ref{superfinalE}) a sufficient condition for the final energy to be smaller than the initial one (i.e. $E_\mathrm{f}<E_\mathrm{i}$) can be obtained,
\begin{eqnarray}
 \frac{2}{3}E_\mathrm{D}+\frac{\hat{l}+1}{3}E_\mathrm{M}<E_\mathrm{D}+E_\mathrm{M}
\end{eqnarray}
\begin{eqnarray}
\Rightarrow \frac{E_\mathrm{D}}{E_\mathrm{M}}> \hat{l}-2.\label{condiwaa}
\end{eqnarray}
This, however, does not ensure that the initial state is unstable, since the energy might not necessarily be monotonically reduced along the rotation, but it shows that a state with lower energy can exist, even when a significant fraction of the energy is contained in a higher order multipole. This is particularly true for the quadrupole, in which case eq. (\ref{condiwaa}) shows that independent of the strength of the quadrupole with respect to the dipole, the final state still has lower energy.

The condition given in eq. (\ref{condiwaa}) might look very restrictive for multipoles with a very high $\hat{l}$, but this is only because it is a sufficient but not a necessary condition.
\subsection{Validity of the assumption of sphericity}\label{far::sphere}
Although the overall change in the structure of the star due to the presence of the magnetic field is small, it is not obvious that the energy associated with this perturbation is smaller than that released by the instability just discussed, and can therefore be ignored. To check if this could in a way stabilise the star against the Flowers-Ruderman instability, we consider the same displacement, where each half is rotated with respect to the other. As the magnetic energy inside the star and the internal energy of the fluid will remain the same, we need only to consider the changes in the external magnetic energy, and the gravitational potential energy.

The change in the structure of the star due to the presence of the magnetic field is related to the plasma $\beta$ parameter (which in our case is $\beta\gg 1$). For instance, the displacement of the surface of the star compared to the perfectly spherical one is of order $\beta^{-1} R$. The change in the external field between both situations is due to the small perturbation of the surface, which scales like $\beta^{-1}$, so, if we use $\vec{B}$ to denote the field in the case where the perturbation to sphericity is ignored, and $\vec{B}+\Delta\vec{B}$ to denote the case where the perturbation to sphericity is considered, we should have that $\Delta \vec{B}/\vec{B}\sim\beta^{-1}$. The energy associated with the external field, if we were to consider the change in the structure of the star, would only be a correction of order $\beta^{-1} E_\mathrm{e}$, so, it can be safely ignored.

We require a slightly more intricate argument to show that the change in the gravitational potential energy during the rotation can be ignored. Consider the star as composed of four different mass distributions, by considering each half separately, and furthermore, by taking each half as composed of a spherical distribution of mass plus a small perturbation due to the presence of the magnetic field. The spherical distribution would have a density $\rho$, while the perturbation $\Delta \rho$ would be of order $\beta^{-1} \rho$. The total gravitational potential energy of the system can then be decomposed in several terms, four corresponding to the individual gravitational energy of each mass distribution, and twelve more related to the interactions of each mass distribution with the other three. Of course, when performing the rotation of one half of the star with respect to the other, the gravitational energy of each of the four elements remains the same, and the only possible change is due to the interactions between them, which can be treated case by case, as follows:
\begin{enumerate}
 \item First, we take the potential energy due to the interaction among the spherical mass distributions. When one half rotates with respect to the other, the system is equivalent to the initial one, so this does not produce a change in the potential energy of the system.
 \item Next, we consider the interaction between a spherical mass distribution and the perturbation of a spherical mass distribution. In this case, a rotation of one half leaves the system composed of these two parts in a state that is equivalent to the initial one, and differs only by a rotation of the whole system. Thus, the gravitational energy in this case remains the same.
 \item Finally, we take the interaction between the two perturbations to the spherical mass distributions. In this case, after the rotation, the system is generally not equivalent to the initial one. The mass associated with each distribution is $\sim \beta^{-1} M$, so the potential energy associated with their interaction is of order
\begin{eqnarray}
  E_g\sim \beta^{-2}\frac{G M^2}{R}=\beta^{-1}\frac{B^2}{P}\frac{G M^2}{R},
\end{eqnarray}
and, since $P\sim GM^2/R^4$,
\begin{eqnarray}
 E_g\sim \beta^{-1} B^2R^3\sim \beta^{-1} E_\mathrm{e},
\end{eqnarray}
so the change in the gravitational potential can in fact be ignored compared with the change in the external magnetic energy. Of course, an exact solid body rotation of each half is not really expected to happen when the star is not spherical, since that would produce some ``rough edges''. But, since the perturbation of sphericity can be ignored in that case, we expect a more reasonable displacement that closely resembles the rotation of each half with respect to the other to exist and reduce the energy even further.
\end{enumerate}
\section{Proof of the instability using perturbation theory}\label{far::pertu}
Using MHD perturbation theory, we should also be able to prove the existence of the Flowers-Ruderman instability. This proof however could not be as complete as the one given in \S\ref{far::proofEx}, since perturbation theory can only be used to see if the system is unstable against small displacements, and thus, we cannot prove with this approach that the energy decreases monotonically along the entire rotation. Nevertheless, we now provide a proof of the instability for a particular family of fields using perturbation theory, since the results obtained in doing so will be useful in the next Section, where we will require the result for the energy perturbation in terms of the angle $\Theta$ by which both halves of the star are rotated.

In \S\ref{far::cvol}, we prove that the volume contribution to the potential energy perturbation is equal to zero. In \S\ref{far::csur}, we calculate the contribution to the potential energy perturbation due to surface currents and show that the final result directly relates with the energy given in eq. (\ref{energybz}). 
\subsection{Contribution to the potential energy perturbation inside the star}\label{far::cvol}
Using the energy principle of \citet{ber+58}, the stability of a system perturbed by a displacement field $\vec{\xi}$ is given by the sign of the potential energy perturbation, which can be written as a sum of hydrostatic and magnetic terms
\begin{eqnarray}
\begin{aligned}
\delta W=&\D\delta W_{\mathrm{hyd}}+\delta W_{\mathrm{mag}},\\
\delta W_{\mathrm{hyd}}=&\D\frac{1}{2}\int_V\left[\Gamma_1 P(\nabla\cdot\vec{\xi})^2+(\vec{\xi}\cdot\nabla P)(\nabla\cdot\vec{\xi})\right.\\
&\qquad\left.-(\vec{\xi}\cdot\nabla\Phi)(\nabla\cdot \rho\vec{\xi})+\rho\vec{\xi}\cdot\nabla\delta\Phi\right]\dd V\\
&\qquad\qquad\D-\frac{1}{2}\oint_S\left(\Gamma_1 P\nabla\cdot\vec{\xi}+\vec{\xi}\cdot \nabla P\right)\vec{\xi}\cdot \dd \vec{s},\\
 \delta W_{\mathrm{mag}}=&\D-\frac{1}{2}\int_V \vec{\xi}\cdot(\delta\vec{j}\times\vec{B}+\vec{j}\times\delta\vec{B})\dd V\label{volwork}
\end{aligned}
\end{eqnarray}
where $V$ now denotes the volume of the star, $P$ is the fluid pressure, $\rho$ is the mass density, $\Phi$ is the gravitational potential, $\Gamma_1$ is defined as
\begin{eqnarray}
 \Gamma_1=\left(\frac{\pp \ln P}{\pp\ln \rho}\right)_{\mathrm{ad}},
\end{eqnarray}
and $\vec{j}$ is the current density, which in a static configuration can be written as
\begin{eqnarray}
 \frac{4\pi}{c}\vec{j}=\nabla\times\vec{B}.
\end{eqnarray}
The magnetic field and current perturbations are given by
\begin{eqnarray}
 \delta\vec{B}=\nabla\times(\vec{\xi}\times\vec{B}),\;\;\frac{4\pi}{c}\delta\vec{j}=\nabla\times\delta\vec{B}.
\end{eqnarray}
If $\delta W<0$, then the resulting configuration will be unstable. We ignore the effects of the magnetic field on the structure of the star, so $P$, $\rho$ and $\Phi$ are spherically symmetric.

The displacement field for the case of the Flowers-Ruderman instability is taken to be
\begin{eqnarray}
 \vec{\xi}=
\begin{cases}
 \;\;\;\Theta r \bhat{x}\times\bhat{r}=-\Theta r (\cos\theta\cos\phi\bhat{\phi}+\sin\phi\bhat{\theta}) & x>0\\
-\Theta r \bhat{x}\times\bhat{r}=\;\;\;\Theta r (\cos\theta\cos\phi\bhat{\phi}+\sin\phi\bhat{\theta}) & x<0
\end{cases}\label{xiFR}
\end{eqnarray}
with $|\Theta|\ll 1$. This displacement field has no radial component and is incompressible, so $\delta W_{\mathrm{hyd}}=0$.

\begin{figure}
\centering
\includegraphics[height=2.5in]{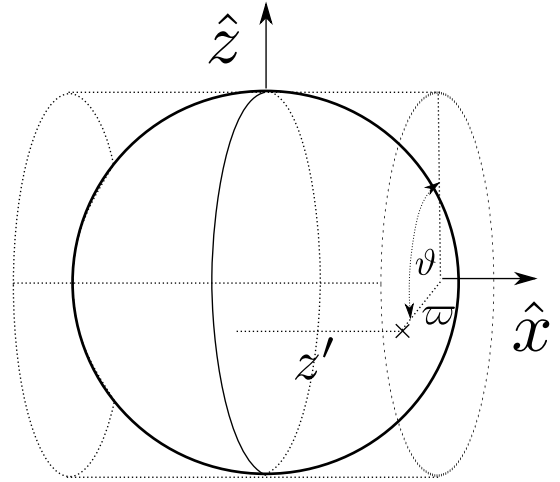}
\caption{Cylindrical coordinate system used in \S\ref{far::cvol}. The coordinates of this system are $\varpi$, $\vartheta$ and $z'$, and the cylinder is oriented in such a way that $\bhat{z}'=\bhat{x}$ and that for $\vartheta=0$ one has $\bhat{\varpi}=\bhat{z}$. Also, the basis vectors of this coordinate system satisfy the relation $\bhat{z}'=\bhat{\varpi}\times\bhat{\vartheta}$, which defines the direction of increasing $\vartheta$.}
\label{cylinderap}
\end{figure}

It can also be proved that for this displacement field, $\delta W_{\mathrm{mag}}=0$ in the bulk of the star, independent of the form of the magnetic field. This is to be expected, since the magnetic field will be simply displaced, and its energy should not change. To do this, we consider a cylindrical coordinate system with coordinates $\varpi$ for the cylindrical radial coordinate, $\vartheta$ for the azimuthal angle and $z'$, as illustrated in Fig. \ref{cylinderap}. In each half of the star, the displacement field can be written as
\begin{eqnarray}
  \vec{\xi}=\pm\Theta\varpi \bhat{\vartheta},
\end{eqnarray}
and in this system, a general magnetic field can be written as
\begin{eqnarray}
  \vec{B}=B_\varpi \bhat{\varpi}+B_\vartheta\bhat{\vartheta}+B_{z'}\bhat{z'},
\end{eqnarray}
where $B_\varpi$, $B_\vartheta$ and $B_{z'}$ are all $2\pi$-periodic functions of $\vartheta$ and must be such that $\nabla\cdot\vec{B}=0$, i.e.
\begin{eqnarray}
  \frac{1}{\varpi}\frac{\pp}{\pp\varpi}(\varpi B_\varpi)+\frac{1}{\varpi}\frac{\pp B_\vartheta}{\pp\vartheta}+\frac{\pp B_{z'}}{\pp z'}=0.
\end{eqnarray}
Using this, the integrand in eq. (\ref{volwork}) can be written as
\begin{eqnarray}
  \vec{\xi}\cdot(\delta\vec{j}\times\vec{B}+\vec{j}\times\delta\vec{B})=\pm\frac{\Theta^2}{4\pi}\frac{\pp F}{\pp \vartheta},
\end{eqnarray}
where $F$ is a $2\pi$-periodic function of $\vartheta$ given by
\begin{eqnarray}
\begin{aligned}
  F=B_{z'}&\left(\frac{\pp B_{z'}}{\pp \vartheta}-\varpi \frac{\pp B_{\vartheta}}{\pp z'}\right)\\
  &\hspace{0.6in}+B_\varpi\left(\frac{\pp B_{\varpi}}{\pp \vartheta}-B_\vartheta-\varpi\frac{\pp B_{\vartheta}}{\pp \varpi}\right).
\end{aligned}
\end{eqnarray}
Since the range of integration for $\vartheta$ in eq. (\ref{volwork}) is from $0$ to $2\pi$, and the integration is done over the derivative of a $2\pi$-periodic function, $\delta W_{\mathrm{mag}}=0$ in the bulk of the star.

However, we already saw that there is an effective variation of the energy when performing this perturbation, and thus, we are not taking into account all the work that is done on the fluid. This large scale displacement produces surface currents in two different regions, and these should be responsible for the work done:
\begin{enumerate}
 \item Along the surface of the sphere. Since the exterior field satisfies Laplace's equation, and its boundary conditions only require the normal component of $\vec{B}$ to be continuous, it is unlikely that a large-scale displacement that affects the surface of the star will not produce a discontinuity of the tangential component of $\vec{B}$ in some areas. Also, even before the rotation is made, surface currents will be present if the field is discontinous along the surface of the star. Thus, surface currents are an important element for perturbations that affect the surface.
 \item Along the plane that cuts the star. The discontinuity produced by the rotation will produce a current sheet along this plane.
\end{enumerate}
From these two effects, only the first is really relevant to the energy of the star. The second effect is not, because $\vec{\xi},\vec{B}$, and $\vec{j}$ are parallel to that surface, and thus no work is done on the fluid. 

\subsection{Contribution to the potential energy perturbation due to surface currents}\label{far::csur}
If a discontinuity of the $\theta$ and $\phi$ components of the magnetic field exists at the surface, a surface current $\vec{K}$ will be produced (see, for instance, \citealt{jac+98}),
\begin{eqnarray}
 \frac{4\pi}{c} \vec{K}=\bhat{r}\times (\vec{B}_{\mathrm{ext}}-\vec{B}_{\mathrm{int}}),
\end{eqnarray}
where $\vec{B}_{\mathrm{int}}$ and $\vec{B}_{\mathrm{ext}}$ are the fields immediately inside and immediately outside the surface of the star respectively. If the field is perturbed by a displacement $\vec{\xi}$, then $\vec{B}_{\mathrm{int}}$ changes to first order in $\vec{\xi}$ by $\delta \vec{B}_{\mathrm{int}}=\nabla\times(\vec{\xi}\times\vec{B}_{\mathrm{int}})$. This change will modify the boundary conditions on the stellar surface, giving rise to a perturbation of the exterior magnetic field
\begin{eqnarray}
 \delta \vec{B}_{\mathrm{ext}}=\nabla \delta \Psi,
\end{eqnarray}
with
\begin{eqnarray}
\begin{aligned}
\delta\Psi=&-\sum_{l,m}\frac{R^{l+2}\delta c_{lm}}{r^{l+1}(l+1)}Y_{lm}(\theta,\phi),\\
\delta c_{lm}=&\int_{4\pi}Y_{lm}^*(\theta,\phi)(\delta B_r)_{r=R}\dd\Omega, \label{deltaphi}
\end{aligned}
\end{eqnarray}
where $\delta B_r=\delta\vec{B}\cdot\bhat{r}$ is the perturbation of the radial magnetic field component. Almost certainly, this will give rise to a perturbation of the surface current:
\begin{eqnarray}
 \frac{4\pi}{c} \delta\vec{K}=\bhat{r}\times (\delta\vec{B}_{\mathrm{ext}}-\delta\vec{B}_{\mathrm{int}}).
\end{eqnarray}
Now, by replacing $\vec{j}$ by $\vec{j}+\delta(r-R)\vec{K}$ in eq. (\ref{volwork}) and performing the radial integral for the term with the unperturbed surface current and the one with the surface current perturbation, the contribution to $\delta W$ due to these terms can be written as\footnote{Considering this, $\delta W_{\mathrm{mag}}$ now consists of a volume integral and a surface integral:
\begin{eqnarray}
\begin{aligned}
\delta W_{\mathrm{mag}}=&-\frac{1}{2}\int_V \vec{\xi}\cdot(\delta\vec{j}\times\vec{B}+\vec{j}\times\delta\vec{B})\dd V\\
&\qquad-\frac{R^2}{2}\int_{4\pi}\left[\vec{\xi}\cdot(\delta \vec{K}\times \vec{B}+\vec{K}\times\delta \vec{B})\right]_{r=R}\dd\Omega.
\end{aligned}
\end{eqnarray}
}
\begin{eqnarray}
 \delta W_{\mathrm{sc}}=-\frac{R^2}{2}\int_{4\pi}\left[\vec{\xi}\cdot(\delta \vec{K}\times \vec{B}+\vec{K}\times\delta \vec{B})\right]_{r=R}\dd\Omega\label{wsc}.
\end{eqnarray}
However, due to the discontinuity of the tangential components of $\vec{B}$ and $\delta\vec{B}$ across the boundary, the choice for these two vectors is somewhat ambiguous, a consequence of the unphysical nature of surface currents. This can be avoided by considering only perturbations that are parallel to the surface, so $\xi_r(R,\theta,\phi)=0$, in which case only the radial components of $\vec{B}$ and $\delta\vec{B}$ contribute to the previous expression, which reduces to
\begin{eqnarray}
\begin{aligned}
\delta W_{\mathrm{sc}}=&\D-\frac{R^2}{8\pi}\left[\int_{4\pi} B_r\vec{\xi}\cdot(\delta\vec{B}_{\mathrm{ext}}-\delta\vec{B}_{\mathrm{int}})\dd\Omega\right.\\
&\D\qquad\qquad\left.+\int_{4\pi}\delta B_r\vec{\xi}\cdot(\vec{B}_{\mathrm{ext}}-\vec{B}_{\mathrm{int}})\dd\Omega\right].\label{wesece} 
\end{aligned}
\end{eqnarray}
Here, it is not necessary to distinguish between the interior and exterior values of $B_r$ and $\delta B_r$ because these must be continuous. The primary difficulty in this expression is the term $\delta\vec{B}_{\mathrm{ext}}$. However, this integral can be explicitly computed in terms of the $\delta c_{lm}$. To do this, we write $\vec{\xi}=\xi_\phi \bhat{\phi}+\xi_\theta \bhat{\theta}$, $\vec{B}=B_r \bhat{r}+B_\theta \bhat{\theta}$, and using integration by parts it can be seen that
\begin{eqnarray}
-\frac{R^2}{8\pi}\int_{4\pi} B_r\vec{\xi}\cdot\delta\vec{B}_{\mathrm{ext}}\dd\Omega=\frac{R^3}{8\pi}\sum_{lm}\frac{|\delta c_{lm}|^2}{l+1},\label{yay}
\end{eqnarray}
so this term is always positive, and thus does not drive the instability.

Now we consider the perturbation field given by eq. (\ref{xiFR}) and a dipolar magnetic field as given by eq. (\ref{brdipole}). In this case, the $\delta c_{lm}$ are
\begin{eqnarray}
 \delta c_{lm}=
\begin{cases}
\D2B_0\Theta w_{lm} & l,m\;\mathrm{even}\\
0 &\mathrm{otherwise}
\end{cases}
\end{eqnarray}
where $B_0$ is the strength of the field at the poles. Using this, together with eq. (\ref{yay}), we get
\begin{eqnarray}
-\frac{R^2}{8\pi}\int_{4\pi} B_r\vec{\xi}\cdot\delta\vec{B}_{\mathrm{ext}}\dd\Omega=A E_\mathrm{e} \Theta^2,\label{equ1}
\end{eqnarray}
where $A$ is given by eq. (\ref{Adef}) and $E_\mathrm{e}$ is given by eq. (\ref{E0def}). This gives us one of the terms of $\delta W_{\mathrm{sc}}$ (as shown in eq. (\ref{wesece})). The other terms can be evaluated directly by using the displacement field given by eq. (\ref{xiFR}) and a magnetic field given by eq. (\ref{brdipole}), yielding the result
\begin{eqnarray}
\begin{aligned}
 &-\frac{R^2}{8\pi}\left[-\int_{4\pi} B_r\vec{\xi}\cdot\delta\vec{B}_{\mathrm{int}}\dd\Omega\right.\\
&\qquad\qquad\quad\left.+\int_{4\pi}\delta B_r\vec{\xi}\cdot(\vec{B}_{\mathrm{ext}}-\vec{B}_{\mathrm{int}})\dd\Omega\right]=-E_\mathrm{e}\Theta^2.\label{equ2}
\end{aligned}
\end{eqnarray}
Considering eqs. (\ref{wesece}), (\ref{equ1}) and (\ref{equ2}), the potential energy perturbation due to the surface currents is found to be
\begin{eqnarray}
 \delta W_{\mathrm{sc}}=(A-1)E_\mathrm{e}\Theta^2
\label{worksc},
\end{eqnarray}
which agrees with eq. (\ref{energybz}) up to order $\Theta^2$, as expected.
\section{Stabilising effect of a toroidal field}\label{csfar}
\begin{figure}
\centering
\includegraphics[height=2in]{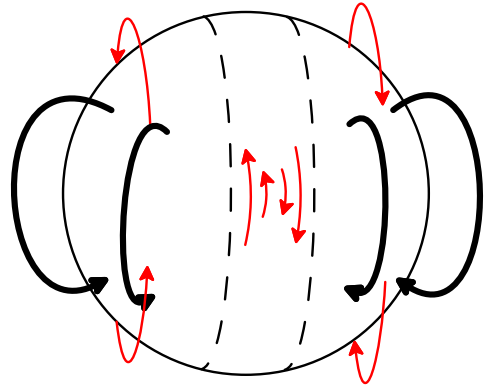}
\caption{Smooth rotation of each half of the star due to the inclusion of a toroidal field. The thick arrows are field lines from the dipole, and the thin lines indicate the rotation of each half of the star. The dashed lines enclose the region of width $2\epsilon R$ where the displacement field switches the direction of rotation continuously, as is shown by the thin arrows there.}
\label{smoothfig}
\end{figure}

When a toroidal field is added, magnetic field lines will pass through the plane that cuts the star in half in the Flowers-Ruderman instability. Thus, if a sharp cut is done, magnetic field lines would be cut, which is not possible. Because of this, an arbitrarily weak toroidal field is enough to stabilise the star against the sharp cut, but if the cut is done smoothly, toroidal field lines will not be cut, but instead will be severely twisted. To study this, we consider the effects of performing the cut of the star smoothly across a region of finite width $2\epsilon R$ (as shown in Fig. \ref{smoothfig}). As $\epsilon$ increases, this bending will be less pronounced, and thus the stabilising effect of the toroidal field will be reduced. Under some reasonable assumptions, we use perturbation theory to obtain a ratio between the energy of the poloidal field and the total energy of the magnetic field for which the field becomes stable to this displacement. This value can be compared with the values obtained by \citet{bra+09} for which the field becomes unstable in MHD simulations.

To do this, we consider a displacement field of the form
\begin{eqnarray}
\vec{\xi}=
\begin{cases}
-\Theta_0 r\bhat{x}\times\bhat{r} & x<-\epsilon R\\
\Theta(x) r\bhat{x}\times\bhat{r} & |x|<\epsilon R\\
\Theta_0 r\bhat{x}\times\bhat{r} & x>\epsilon R
\end{cases}\label{xiSFR}
\end{eqnarray}
where $\Theta(x)$ is a continuous, odd function in the interval $|x|<\epsilon R$ that satisfies
\begin{eqnarray}
\begin{aligned}
\Theta(\pm \epsilon R)=&\pm \Theta_0,\;\;\frac{d \Theta}{d x}_{x=\pm \epsilon R}=0.
\end{aligned}\label{condi}
\end{eqnarray}
The condition imposed on the derivative is to avoid discontinuities in $\delta \vec{B}$ along the boundary, which would in turn produce surface currents. Similar to the displacement field used before for the sharp cut, $\vec{\xi}$ has no $\bhat{r}$ component and satisfies $\nabla\cdot\vec{\xi}=0$, so there will be no hydrostatic contribution to $\delta W$ as can be seen from eq. (\ref{volwork}).

The potential energy perturbation for this displacement field can be split into several terms, including a term that involves surface currents, $\delta W_{\mathrm{sc}}$. This contribution to the potential energy perturbation involves surface integrals of an infinite number of spherical harmonics, and the fact that the displacement field is defined in terms of Cartesian coordinates adds great complexity in trying to evaluate $\delta W_{\mathrm{sc}}$. Because of this, we consider that the smooth transition is done in a thin region relative to the radius of the star, so $\epsilon\ll 1$, and we assume that $\delta W_{\mathrm{sc}}$ does not change significantly with respect to the value obtained for the sharp cut\footnote{We do not expect the external magnetic field to be significantly different on the surface of the star for the region $|x|>\epsilon R$, so the contribution to $\delta W_{\mathrm{sc}}$ on this region should not change significantly. Also, the area of the surface in the region $|x|<\epsilon R$ is small compared to the rest of the surface in which the integral for $\delta W_{\mathrm{sc}}$ is done, so even if there are significant changes there, we do not expect them to significantly modify the work done on the whole surface.}. In any case, we expect $\delta W_{\mathrm{sc}}$ to increase as $\epsilon$ increases, since in this case the dipole component of the external magnetic field will not be reduced as much as in the case of the sharp cut.

\subsection{Structure of the toroidal field}\label{apii}
In an axisymmetric hydromagnetic equilibrium, the magnetic field and its associated currents cannot produce forces in the azimuthal direction, because the pressure and gravity forces would not be able to balance them. This implies that the toroidal magnetic field is contained in a particular region of the star that is restricted by the topology of the poloidal field, as will be shown in this section.

The most general axially symmetric magnetic field can be decomposed into a toroidal and a poloidal part, each of which is determined by a scalar function (see, for instance, \citealt{chapre+56}),
\begin{eqnarray}
 \vec{B}=\vec{B}_{\mathrm{T}}+\vec{B}_{\mathrm{P}},\quad \vec{B}_{\mathrm{T}}=\beta \nabla\phi,\quad \vec{B}_{\mathrm{P}}=\nabla \alpha \times \nabla \phi \label{equap}
\end{eqnarray}
where\footnote{In this subsection we use $\beta$ to represent the function that defines the toroidal field, instead of its usual meaning as a ratio of fluid to magnetic pressure.} $\beta=\beta(r,\theta)$ and $\alpha=\alpha(r,\theta)$. Since $\vec{B}_{\mathrm{P}}\perp \nabla \alpha$ and $\vec{B}_{\mathrm{T}}\perp \nabla\beta$, $\alpha$ and $\beta$ are constant along poloidal and toroidal field lines respectively. The force per unit volume exerted by the magnetic field is given by
\begin{eqnarray}
\begin{aligned}
 c\vec{F}_\mathrm{M}=&\vec{j}\times\vec{B}\\
=&\vec{j}_{\mathrm{P}}\times\vec{B}_{\mathrm{P}}+\vec{j}_{\mathrm{P}}\times\vec{B}_{\mathrm{T}}+\vec{j}_{\mathrm{T}}\times\vec{B}_{\mathrm{P}}+\vec{j}_{\mathrm{T}}\times\vec{B}_{\mathrm{T}},
\end{aligned}
\end{eqnarray}
where $\vec{j}_{\mathrm{P}}$ and $\vec{j}_{\mathrm{T}}$ are the currents related to the poloidal and toroidal fields,
\begin{eqnarray}
 \frac{4\pi}{c}\vec{j}_{\mathrm{P}}=\nabla\times \vec{B}_{\mathrm{P}},\qquad \frac{4\pi}{c}\vec{j}_{\mathrm{T}}=\nabla\times \vec{B}_{\mathrm{T}}.
\end{eqnarray}
Since $\vec{j}_{\mathrm{P}}$ and $\vec{B}_{\mathrm{T}}$ are both toroidal fields, their cross product is zero. Also, $\vec{j}_{\mathrm{P}}\times \vec{B}_{\mathrm{P}}$ and $\vec{j}_{\mathrm{T}}\times \vec{B}_{\mathrm{T}}$ are both the cross products of a toroidal and a poloidal field, so they have no $\bhat{\phi}$ component. The remaining term, $\vec{j}_{\mathrm{T}}\times\vec{B}_{\mathrm{P}}$ is the cross product of two poloidal fields, so it points in the $\bhat{\phi}$ direction, and the $\bhat{\phi}$ component of the magnetic force is given by
\begin{eqnarray}
 cF_{M\phi}\bhat{\phi}=\vec{j}_{\mathrm{T}}\times\vec{B}_{\mathrm{P}}.
\end{eqnarray}
If the configuration is in axisymmetric equilibrium, $F_{M\phi}$ must vanish since there is no possible way for the fluid to counteract this magnetic force, so $\vec{j}_{\mathrm{T}}\parallel\vec{B}_{\mathrm{P}}$. The current $\vec{j}_{\mathrm{T}}$ can be calculated as
\begin{eqnarray}
 4 \pi \vec{j}_{\mathrm{T}}=\nabla\times \vec{B}_{\mathrm{T}}=\nabla\beta\times\nabla\phi,
\end{eqnarray}
so the condition that $\vec{j}_{\mathrm{T}}\parallel\vec{B}_{\mathrm{P}}$ is equivalent to the condition $(\nabla\beta\times\nabla\phi)\parallel(\nabla\alpha\times\nabla\phi)$, and since $\nabla\beta$ and $\nabla\alpha$ have no $\bhat{\phi}$ component, $\nabla\beta\parallel\nabla\alpha$ which means that $\alpha$ and $\beta$ can be written as functions of each other, $\alpha=\alpha(\beta)$ or $\beta=\beta(\alpha)$. This result was originally derived by \citet{chapre+56}.

Now, consider a poloidal field line that closes outside a star that is in equilibrium. Along this field line, $\alpha$ is constant, and thus, $\beta=\beta(\alpha)$ is also a constant. However, outside the star the field cannot have a toroidal component, so $\beta=0$ in that region, and since $\beta$ is constant along the field line, the toroidal field vanishes everywhere along a poloidal field line that closes outside the star. Because of this, the toroidal field must be contained in regions where the poloidal field lines are closed within the star. This is not a new result, since this condition for axisymmetric hydromagnetic equilibria has been known for many years (e.g. \citealt{rob+81}).	

In Fig. \ref{poloi} we plot the field lines of the particular poloidal component of the magnetic field we use in this section (this field is described in greater detail in \S\ref{poloisec}). For this poloidal field, the toroidal field is enclosed in a torus-like region, so over most of the star the field is purely poloidal.

\subsection{Cylinder approximation and toroidal fields}\label{cylsection}
As a simple approximation to the region of transition ($|x|<\epsilon R$), we will consider it as a cylinder of height $2\epsilon R$ and radius $R$, and use the same cylindrical coordinate system used in \S\ref{far::cvol}.

The displacement field of eq. (\ref{xiSFR}) in this region can be written as
\begin{eqnarray}
\vec{\xi}=\Theta(z')\varpi\bhat{\vartheta}.
\end{eqnarray}
We consider the perturbation in the potential energy of a toroidal field due to this displacement. Since the height of the cylinder is small relative to the radius of the star, we approximate the toroidal field as
\begin{eqnarray}
\vec{B}_{\mathrm{T}}=b(\varpi,\vartheta)\bhat{z}'
\end{eqnarray}
where $b(\varpi,\vartheta)$ is a $2\pi$-periodic function that is odd in $\vartheta$ (i.e. $b(\varpi,-\vartheta)=-b(\varpi,\vartheta)$). Using this, the potential energy perturbation in this region due solely to the toroidal field can be obtained from eq. (\ref{volwork}),
\begin{eqnarray}
\begin{aligned}
\delta W_{\mathrm{T}}=&\D-\frac{1}{8\pi}\int_{-\epsilon R}^{\epsilon R}\dd z'\int_0^{2\pi}\dd\vartheta\int_{0}^{R}\dd\varpi \\&\D\qquad\qquad\left[\Theta^2 \frac{\pp}{\pp \vartheta}\left(b\frac{\pp b}{\pp\vartheta}\right)+\varpi^2 b^2 \Theta \frac{d^2 \Theta}{d z'^{2}}\right]\varpi. 
\end{aligned}
\end{eqnarray}
The first term vanishes after integration over $\vartheta$, and since we demand that the derivative of $\Theta$ vanishes on the boundary, the second term can be rewritten after integration by parts as
\begin{eqnarray}
\delta W_{\mathrm{T}}=\frac{1}{8\pi}\int_{-\epsilon R}^{\epsilon R}\dd z' \left(\frac{d \Theta}{d z'}\right)^2\int_0^{2\pi}\dd\vartheta\int_{0}^{R}\dd\varpi\; \varpi^3 b^2.
\end{eqnarray}
From this, it can be seen immediately that $\delta W_{\mathrm{T}}>0$, so, as expected, the toroidal field opposes this displacement. 

We now need to specify a model for both $\Theta(z')$ and $b(\varpi,\vartheta)$. We choose our function $\Theta(z')$ as
\begin{eqnarray}
\begin{aligned}
 \Theta(z')=&\D\Theta_0\sin\left(\frac{\pi z'}{2\epsilon R}\right)\\
\Rightarrow\quad&\D \delta W_{\mathrm{T}}=\frac{\pi \Theta_0^2}{32\epsilon R}\int_0^{2\pi}\dd\vartheta\int_{0}^{R}\dd\varpi \;\varpi^3 b^2.\label{Omegadef}
\end{aligned}
\end{eqnarray}
This function $\Theta(z')$ is odd and satisfies the required conditions mentioned in eq. (\ref{condi}). With this particular displacement field, $\delta W_{\mathrm{T}}\propto \epsilon^{-1}$, so, as mentioned before, if the region where the displacement field switches direction is very thin, the magnetic energy will increase significantly, and thus an infinitely weak toroidal field is enough to stabilise the star against a sharp cut.

\begin{figure}
\centering
\includegraphics[height=2.4in]{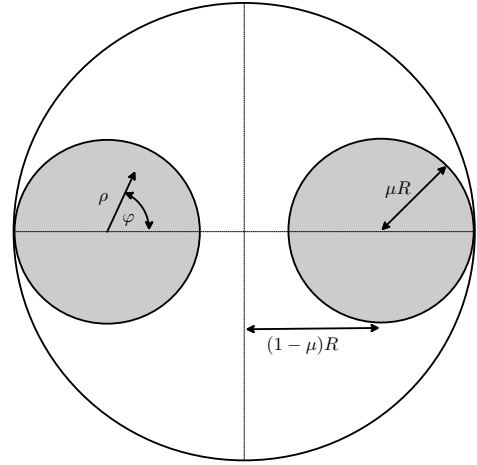}
\caption{Model used for the toroidal field. The vertical line is the symmetry axis, and the field is contained in a circular torus of radius $\mu R$ that is tangent to the equator of the star, as shown by the shaded region in the figure. Also depicted in the figure are the coordinates $\rho$ and $\varphi$ used to describe the magnitude of the field. In the cylinder approximation the torus is treated as two independent cylindrical regions.}
\label{torField}
\end{figure}

Since the toroidal field is confined within the poloidal field lines that are closed inside the star (as is shown in \S\ref{apii}), we consider the toroidal field to be contained in a torus of internal radius $\mu R$. This approximation is adequate for the particular poloidal field we will use, as can be seen in Fig. \ref{poloi}. In the cylinder approximation, we consider this torus as two cylindrical regions of radius $\mu R$ that are centred at $(\varpi,\vartheta)=(R(1-\mu),\pm \pi/2)$, as illustrated in Fig. \ref{torField}. In each of these regions, the strength of the field will depend on the distance to the centre, so we switch to coordinates $(\rho,\varphi)$ centred on one of these circles in which we have $b=b(\rho)$ (as is shown in Fig. \ref{torField}). The corresponding $\delta W_{\mathrm{T}}$ can be solved in these coordinates as
\begin{eqnarray}
\delta W_{\mathrm{T}}=\D\frac{\pi \Theta_0^2}{16\epsilon R}\int_0^{2\pi}\dd\varphi\int_{0}^{\mu R}\dd\rho \;\rho b^2(\rho)d^2(\rho,\varphi)\label{wtorwaaa},
\end{eqnarray}
where $d^2(\rho,\varphi)=\rho^2+[R(1-\mu)]^2-2\rho R(1-\mu)\cos \varphi$ is the distance to the origin. As a model for $b(\rho)$, we use
\begin{eqnarray}
b(\rho)=\eta B_0\cos^2\left(\frac{\rho \pi}{2 \mu R}\right),
\end{eqnarray}
where $B_0$ is the maximum strength of the poloidal field on the surface, and $\eta B_0$ is the maximum strength of the toroidal field. The square on the cosine assures $\delta \vec{B}$ to be continuous along the surface where the toroidal field vanishes. Using this model for the field, $\delta W_{\mathrm{T}}$ results in
\begin{figure}
\centering
\includegraphics[height=3in,angle=270]{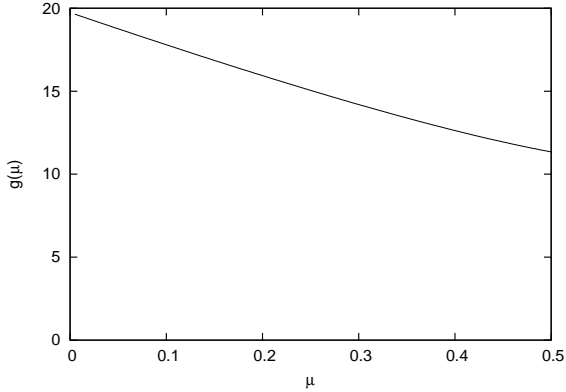}
\caption{Plot of the function $g(\mu)$ from eq. (\ref{wipiii2})}
\label{gmu}
\end{figure}
\begin{eqnarray}
\begin{aligned}
\delta W_{\mathrm{T}}=&\frac{3 E_\mathrm{e} \Theta_0^2}{64 \pi^2}\frac{\eta^2 h(\mu)}{\epsilon}\\
h(\mu)=&\pi^2(6\pi^2-32)\mu^2(1-2\mu)\\
&\hspace{0.6in}+(9\pi^4-77\pi^2+192)\mu^4
,\end{aligned}
\label{wipiii}
\end{eqnarray}
where $E_\mathrm{e}$ is the initial energy of the exterior magnetic field as given by eq. (\ref{E0def}).

It can be seen from eq. (\ref{wtorwaaa}) that the detailed geometry of the toroidal field is not so relevant, specially if the toroidal field is contained in a region far away from the centre of the star. In the latter case, $d(\rho,\varphi)\simeq R(1-\mu)$, and the integral will involve only the square of the magnitude of the magnetic field times an area element. Because of this, $\delta W_{\mathrm{T}}$ should be closely related to the energy of the toroidal field, rather than depend on its detailed geometry. The energy of the toroidal field can be computed as
\begin{eqnarray}
 E_{\mathrm{T}}=\D\frac{3 E_\mathrm{e}}{8 \pi}\left(3\pi^2-16\right)\eta^2\mu^2(1-\mu),
\end{eqnarray}
which can be used to rewrite eq. (\ref{wipiii}) as
\begin{eqnarray}
  \begin{aligned}
    \delta W_\mathrm{T}=&\frac{E_\mathrm{T}\Theta_0^2}{8\pi}\frac{g(\mu)}{\epsilon}\\
    g(\mu)=&\frac{h(\mu)}{\left(3\pi^2-16\right)\mu^2(1-\mu)}.
  \end{aligned}
\label{wipiii2}
\end{eqnarray}
A plot of the function $g(\mu)$ is shown in Fig. \ref{gmu}, where it can be seen that at a fixed energy of the toroidal field, the stabilizing effect will be stronger if the field is contained farther away from the center of the star. Also, since $g(\mu)$ only varies by a factor $\sim 2$, it is the energy of the toroidal field what sets the order of magnitude of $\delta W_\mathrm{T}$, while its geometry plays a secondary role.
\subsection{Effect of poloidal fields for the smooth rotation}
\subsubsection{Cross term in $\delta W$}\label{crosi}
When a poloidal field is added, a cross term appears in $\delta W$ that involves both the poloidal and toroidal components of the magnetic field. This term has the form
\begin{eqnarray}
\begin{aligned}
 \delta W_{\mathrm{cross}}=&-\frac{1}{2}\int_V\dd V\vec{\xi}\cdot\left[\delta \vec{j}_{\mathrm{T}}\times\vec{B}_{\mathrm{P}}+\vec{j}_{\mathrm{T}}\times\delta\vec{B}_{\mathrm{P}}\right.\\
&\left.\qquad\qquad\qquad\qquad+\delta \vec{j}_{\mathrm{P}}\times\vec{B}_{\mathrm{T}}+\vec{j}_{\mathrm{P}}\times\delta\vec{B}_{\mathrm{T}}\right].
\end{aligned}
\end{eqnarray}
Considering only the parity of the functions involved, it can be shown that the integrand in $\delta W_{\mathrm{cross}}$ is an odd function of $x$, and since the integral is over the interval $-\epsilon R<x< \epsilon R$, integration over $x$ will immediately give zero as a final result, so
\begin{eqnarray}
\delta W_{\mathrm{cross}}=0.
\end{eqnarray}

\subsubsection{Purely poloidal contribution to $\delta W$}\label{poloisec}
For the poloidal field, we will consider a configuration of the form
\begin{eqnarray}
\vec{B}=\nabla\alpha\times\nabla\phi,\quad\alpha=f(r)\sin^2\theta. \label{fieldpol}
\end{eqnarray}
On the surface of the star, the radial component for these fields is $\D2f(R)R^{-2}\cos(\theta)$, and thus, outside the star all these fields are pure dipoles. This model for the internal field covers a wide range of axisymmetric configurations; this includes the constant field $\vec{B}=B_0\hat{z}$ in which $f(r)=B_0r^2/2$ and the fields used by \citet{bra+07} to study the stability of purely poloidal fields in rotating stars.

\begin{figure}
\centering
\includegraphics[height=2.8in]{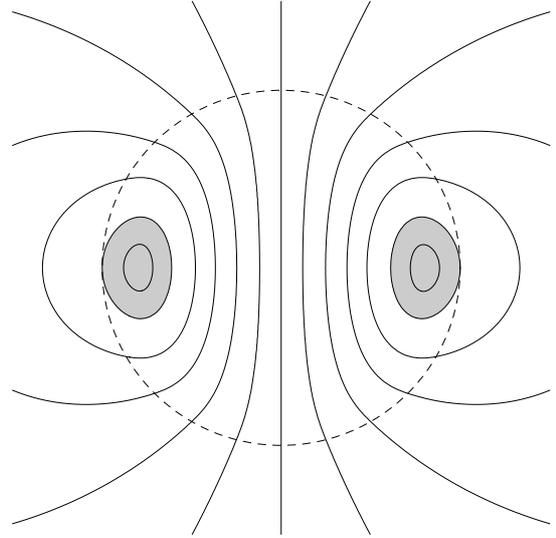}
\caption{Field lines for the particular poloidal field configuration we use in this section. The dashed line represents the surface of the star, the solid lines are the field lines, and the region where poloidal field lines are closed inside the star is marked in grey. This field is of the form given by eq. (\ref{fieldpol}), with $f(r)$ given by eq. (\ref{efepol}). The region where poloidal field lines are closed inside the star has a shape similar to a torus, and the toroidal field must be contained there. Figure adapted with permission from Akg\"un et al. (paper in preparation).}
\label{poloi}
\end{figure}

Our choice for the function $f(r)$ is equivalent to that of Akg\"un et al. (paper in preparation), but normalised so at the poles the strength is $B_0$. This poloidal field is of the form of eq. (\ref{fieldpol}), with
\begin{eqnarray}
f(r)=\frac{35 B_0}{16}\left(r^2-\frac{6}{5}\frac{r^4}{R^2}+\frac{3}{7}\frac{r^6}{R^4}\right).\label{efepol}
\end{eqnarray}
This field is completely continuous across the surface of the star, so there are no surface currents present in the equilibrium configuration. Also, it satisfies $|\vec{j}|=0$ at the surface, which is expected from the fact that the matter density goes to zero there. A plot of this field is shown in Fig. \ref{poloi}.

Using the cylinder approximation, it is difficult to treat the contribution to $\delta W$ due only to the poloidal field. It is also difficult to treat the problem in spherical coordinates, since the regions of integration involved are non-trivial. However, for our particular choice of the poloidal field, the purely poloidal contribution to $\delta W$ can be obtained exactly using Cartesian coordinates. The displacement field used here is of the form of eq. (\ref{xiSFR}), and the function $\Theta(x)$ is the same as that of eq. (\ref{Omegadef}) with $z'$ replaced by $x$.

\begin{figure}
\centering
\includegraphics[height=2.5in]{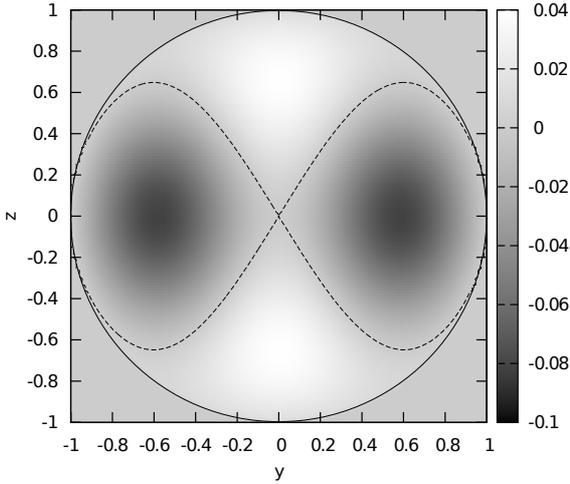}
\caption{Integrand for $\delta W_{\mathrm{P}}$, $-\vec{\xi}\cdot(\delta\vec{j}_{\mathrm{P}}\times\vec{B}_{\mathrm{P}}+\vec{j}_{\mathrm{P}}\times\delta\vec{B}_{\mathrm{P}})/2$, in the plane given by $x=\epsilon/2$ using $\epsilon=1/10$, with the line where the integrand is equal to zero plotted on top. The regions that contribute negatively to $\delta W_{\mathrm{P}}$ (black in the figure) correspond closely to the region where poloidal field lines are closed inside the star, meanwhile the regions that contribute positively to $\delta W_{\mathrm{P}}$ (white in the figure) lie near to the symmetry axis.}
\label{smooth}
\end{figure}

With all this, the potential energy perturbation due solely to the poloidal field can be computed from eq. (\ref{volwork}), which was done using the software Maxima\footnote{\href{http://maxima.sourceforge.net}{http://maxima.sourceforge.net}}. The result is a finite polynomial in $\epsilon$, to lowest order
\begin{eqnarray}
\begin{aligned}
\delta W_{\mathrm{P}}&=\frac{(23\pi^2-330)}{8192}B_0^2 R^3\Theta_0^2\epsilon=\frac{(69\pi^2-990)}{2048}E_\mathrm{e}\Theta_0^2\epsilon\\
&\simeq -0.15 E_\mathrm{e} \Theta_0^2 \epsilon. \label{polwork}
\end{aligned}
\end{eqnarray}
This contribution is negative, but it is not as important as that of $\delta W_{\mathrm{sc}}$ (from eq. (\ref{worksc}) it can be seen that $\delta W_{\mathrm{sc}}\simeq -0.45 E_\mathrm{e} \Theta^2$). Initially we expected the poloidal field to perform a stabilising effect, since this displacement would tend to twist field lines that are near to the symmetry axis. However, the region where the poloidal field lines are closed within the star turns out to be highly unstable to this displacement, as can be seen in Fig. \ref{smooth}. It can be seen that the contribution to the potential energy perturbation is positive along the axis of symmetry, and the region where it is negative encloses the field lines that are closed inside the star. We believe the positive contribution to be caused by the twisting of field lines, and the negative contribution to be due to the displacement of closed poloidal field lines, which resembles a kink mode as described by Markey \& Tayler (1973) and Wright (1973). In a kink mode, closed poloidal field loops inside the star are displaced with respect to each other mostly in a direction perpendicular to the neutral line, but still, the displacement parallel to the neutral line is fundamental. Perhaps adding such component to our analysis could cause an important decrease in $\delta W_{\mathrm{P}}$, but this would significantly complicate our analysis, and it is unlikely that we can find an analytic solution for $\delta W_{\mathrm{P}}$ under those circumstances.
\subsection{Total potential energy perturbation}
To obtain the total energy perturbation, we add all the contributions obtained so far, given by eqs. (\ref{worksc}), (\ref{wipiii2}) and (\ref{polwork}),
\begin{eqnarray}
\delta W=&\hspace{-1.5in}\delta W_{\mathrm{sc}}+\delta W_{\mathrm{T}}+\delta W_{\mathrm{P}}\\
=&\D\hspace{-0.12in} E_\mathrm{e} \Theta_0^2 \left[-(1-A)+\frac{E_\mathrm{T}}{E_\mathrm{e}}\frac{g(\mu)}{8\pi\epsilon}-\frac{990-69\pi^2}{2048}\epsilon\right].
\end{eqnarray}
If $\delta W=0$, then the system is marginally stable, and for that case, solving for $E_{\mathrm{T}}/E_\mathrm{e}$ in terms of $\mu$ and $\epsilon$ results in
\begin{eqnarray}
\frac{E_{\mathrm{T}}}{E_\mathrm{e}}=\frac{8\pi\epsilon}{g(\mu)}\left[(1-A)+\frac{(990-69\pi^2)}{2048}\epsilon\right]\label{etaquad},
\end{eqnarray}
which is an increasing function of both $\mu$ and $\epsilon$ in the range that these variables cover ($0<\mu<0.5$, $0<\epsilon<1$). Choosing $\mu$ and $\epsilon$, we obtain a lower bound on the strength of the toroidal field needed to stabilise the star against a smooth rotation done over a region of width $2\epsilon R$. However, $\mu$ is not completely arbitrary, since in equilibrium, the toroidal field must be contained by the field lines that are closed inside the star. A reasonable value for $\mu$ (for the poloidal field chosen) is $\mu=0.2$, and slight variations from this value do not modify the result significantly.

The choice for $\epsilon$ must be a small value in order for our approximations to be valid. We choose in a rather arbitrary way the value $\epsilon=1/3$ for which the region in which the displacement field switches the direction of rotation has a width of $2R/3$. Under these conditions the assumption that $\delta W_{sc}$ is similar to the one calculated for the straight cut in \S\ref{far::csur} is very doubtful, so this choice of $\epsilon$ certainly serves as an upper bound.

Using $\mu=0.2$, $\epsilon=1/3$ and $A=0.546$ (which is taken from eq. (\ref{Alimits})) to evaluate $E_{\mathrm{T}}/E_\mathrm{e}$ in eq. (\ref{etaquad}), one obtains $E_{\mathrm{T}}/E_\mathrm{e}\sim0.27$.
\subsection{Comparing the poloidal and toroidal energy of the magnetic field}
In order to compare this result with that of Braithwaite (2009), we must see what it means in terms of the energies of the toroidal and poloidal fields (for the latter, including the volume outside the star). The energy of the poloidal field can be evaluated as
\begin{eqnarray}
 E_{\mathrm{P}}=\frac{35}{66}B_0^2R^3=\frac{70}{11}E_\mathrm{e}.
\label{energiesFinal}
\end{eqnarray}
With this, the ratio of poloidal to total energy is
\begin{eqnarray}
\frac{E_{\mathrm{P}}}{E}=\frac{E_{\mathrm{P}}}{E_{\mathrm{T}}+E_{\mathrm{P}}}=\frac{1}{\D\frac{11}{70}\frac{E_\mathrm{T}}{E_\mathrm{e}}+1}\label{eratio}.
\end{eqnarray}
For the value $E_{\mathrm{T}}/E_\mathrm{e}\sim0.27$ obtained in the previous section, this ratio is very close to unity, $E_{\mathrm{P}}/E\sim0.96$. This tells us that a toroidal field with an energy much smaller than the poloidal field is enough to stabilise the star against this perturbation. This can be compared with the instability that could be seen in the simulations by \citet{bra+09} for a ratio of $E_{\mathrm{P}}/E=0.8$. As this perturbation happens with a much stronger toroidal field, it seems to indicate that the perturbation we are studying is not the dominant one, since other instabilities are present for the poloidal field even when the toroidal field is strong enough to stabilise it against the one we have studied.
\section{Effect of successive cuts}\label{far::succesive}
Some authors, for instance, \citet{rob+81} and \citet{branor+06}, have stated that a second cut in a direction perpendicular to the first one would produce a configuration which resembles an octupole. Additionally, they expected that this process could be repeated {\it ad infinitum}, with cuts in different directions, to produce a configuration that resembles an arbitrary multipole. Since the external energy of the field is expected to be reduced as higher order multipoles are achieved, in principle they expected the process to proceed naturally, reaching very high order multipoles, for which the external energy of the field is negligible. \citet{rob+81} computed the energies associated with the external magnetic field for several of these multipoles, showing that the energy of each multipole was in fact smaller than that of the previous one. Even though our result differs slightly from the value obtained by Roberts for the quadrupole, we expect this tendency to be true.

\begin{figure}
\centering
\includegraphics[height=2.1in]{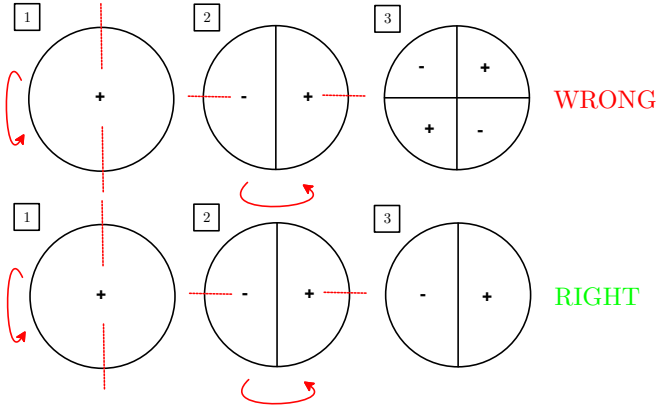}
\caption{Performing two cuts to the dipole. The star is seen from the top of the symmetry axis, and the plus and minus signs indicate magnetic lines coming out and into the star respectively. At the top the wrong picture is shown, where an octupole is produced, while the bottom shows that actually the second cut leaves the star in the dipole configuration.}
\label{twocuts}
\end{figure}

\begin{figure}
\centering
\includegraphics[height=2.1in]{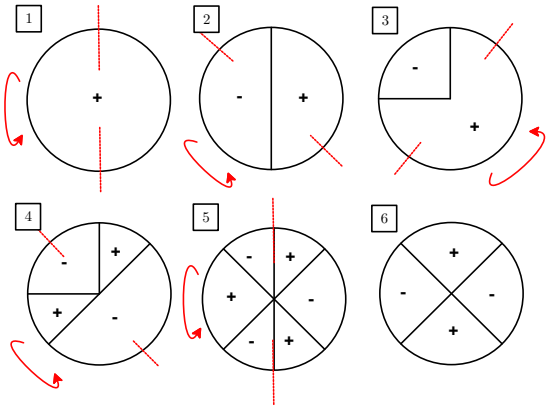}
\caption{Performing several cuts to the dipole in order to obtain an octupole configuration.}
\label{fivecuts}
\end{figure}

However, this process breaks down already at the second cut, as is shown in Fig. \ref{twocuts}, where it is seen that the second cut leaves the star again in the quadrupole configuration, instead of producing an octupole\footnote{Note that the terms quadrupole and octupole used here do not refer to a pure quadrupole or octupole, but to a configuration in which the primary component is that of a quadrupole or an octupole respectively. In effect, these configurations have contributions from an infinite number of multipoles.}.
It is possible to obtain the octupole with a sequence of several cuts, as shown in Fig. \ref{fivecuts}, but through this process the energy does not seem to be monotonically reduced, since intermediate steps have important contributions from a dipole component, so the star cannot actually follow this sequence of displacements.
Even though we only showed a particular sequence of cuts that transforms the quadrupole into an octupole, clearly this cannot be achieved by a single cut, and it does not seem plausible that this can be done without increasing the external magnetic energy at some point, so we expect this mechanism to affect only the initial axisymmetric field.

\section{Conclusions}\label{conclusions}
\citet{florud+77} presented an argument that shows how purely poloidal fields in stars are unstable. If the external field is similar to a dipole, one could cut the star in half and rotate each piece in opposite directions, leading to a configuration in which the external field resembles a quadrupole, and thus, the energy of the external magnetic field should be significantly reduced. Although the Flowers-Ruderman instability is widely accepted, no formal proof had been given that shows both that the external magnetic energy is reduced when the rotation of each half is completed and that the energy decreases monotonically along the entire process.

In this work, we presented a formal proof of this mechanism for the case in which the initial field outside the star is that of a point dipole, by computing the energy of the external field along the entire rotation. We showed that the external magnetic energy decreases monotonically, having a final value of approximately $0.55E_\mathrm{e}$, where $E_\mathrm{e}$ is the initial external magnetic energy as given by eq. (\ref{E0def}).

We also studied the Flowers-Ruderman instability using perturbation theory, in which case we had to consider the effects of surface currents in order for the instability to appear. These effects are not unique to the Flowers-Ruderman instability, and should be considered for any displacement that modifies the magnetic field on the surface. The result obtained for the potential energy perturbation of the star was found to be consistent with the exact value of the energy previously found.
 
We then studied how a toroidal field could stabilise the star against the Flowers-Ruderman instability. Since a sharp cut through the star would split toroidal field lines, the rotation has to be carried out with a continuous displacement field that switches the orientation of rotation across a thin region. For a specific model, it was found that the configuration was stable against the Flowers-Ruderman instability for a ratio of poloidal magnetic energy to total magnetic energy of $E_{\mathrm{P}}/E\lesssim0.96$. Using MHD simulations, \citet{bra+09} had shown that when the ratio $E_{\mathrm{P}}/E$ was below $0.8$, the instabilities driven by the poloidal field were suppressed, but if the ratio was just above $0.8$, the field was found to be unstable with an $m=2$ mode that does not resemble the Flowers-Ruderman instability. Because of this, we conclude that the Flowers-Ruderman instability is not the dominant one.

However, the critical value for the ratio $E_{\mathrm{P}}/E$ we obtained is not a very accurate measure of the stability of the system against the perturbation we treated, since the contribution to the potential energy perturbation comes from different sources which do not necessarily scale with the energy of the poloidal field. If a configuration only has a small region with closed poloidal field lines, then the kinks that produce the negative contribution to the energy described in \S\ref{poloisec} will not be important. In a similar way, the magnetic field can be very strong inside the star, but weak at the surface, so the energy released by the Flowers-Ruderman instability can change by orders of magnitude between configurations with similar poloidal magnetic energy. In the case when the effect due to the closed poloidal fields is not important, a good indicator should be the ratio of external magnetic energy to toroidal magnetic energy, which for our case results in $E_\mathrm{T}/E_\mathrm{e}\simeq0.27$ as the critical value, which is computed from eq. (\ref{etaquad}) with the values previously obtained for $\mu$ and $\eta$.


\section*{Acknowledgments}
This project is supported by FONDECYT Regular Project 1060644, \hspace{0.01in}FONDECYT\hspace{0.01in} Postdoctoral\hspace{0.01in} Project\hspace{0.01in} 3085041, FONDAP Centre for Astrophysics (15010003), Proyecto Basal PFB-06/2007, Proyecto L\'imite VRI 2010-15, Proyecto Alma-Conicyt 3109002 and Conicyt graduate fellowship.


\begin{thebibliography}{}
\bibitem[Arfken \& Weber (2005)]{arf+05} Arfken, G. \& Weber, H., 2005, Mathematical Methods for Physicists, Academic Press
\bibitem[Babcock (1947)]{bab+47} Babcock, H.W., 1947, ApJ, 105, 105
\bibitem[Bernstein et al. (1958)]{ber+58} Bernstein, I.B., Frieman, E.A., Kruskal, M. D. \& Kulsrud, R. M. 1958, Proc. Roy. Soc. A, 244, 17
\bibitem[Braithwaite (2007)]{bra+07} Braithwaite, J., 2007, A\&A, 469, 275
\bibitem[Braithwaite (2009)]{bra+09} Braithwaite, J., 2009, MNRAS, 397, 763
\bibitem[Braithwaite et al.(2010)]{bra+10} Braithwaite, J. et al., 2010, Highlights of Astronomy, Volume 15, p. 161-171
\bibitem[Braithwaite \& Nordlund (2006)]{branor+06} Braithwaite, J. \& Nordlund, A., 2006, A\&A, 450, 1077
\bibitem[Braithwaite \& Spruit (2004)]{braspru+04} Braithwaite, J. \& Spruit, H.C., 2004, Nat, 431, 891
\bibitem[Braithwaite \& Spruit (2006)]{braspru+06} Braithwaite, J. \& Spruit, H.C., 2006, A\&A, 450, 1097
\bibitem[Chandrasekhar \& Prendergast (1956)]{chapre+56} Chandrasekhar, S., \& Prendergast, K. H. 1956, Proc. Nat. Acad. Sci., 42, 5
\bibitem[Flowers \& Ruderman (1977)]{florud+77} Flowers, E. \& Ruderman, M.A., 1977, ApJ, 215, 302
\bibitem[Heger, Woosley \& Spruit (2000)]{hewosp+00} Heger, A., Woosley, S. E. \& Spruit, H.C., 2005, ApJ, 626, 350
\bibitem[Jackson (1998)]{jac+98} Jackson, J., 1998, Classical Electrodynamics, Wiley, New York
\bibitem[Marchant (2010)]{mar+10} Marchant, P., 2010, undergraduate thesis, Pontificia Universidad Cat\'olica de Chile
\bibitem[Markey \& Tayler (1973)]{martay+73} Markey, P. \& Tayler, R.J., 1973, MNRAS, 163, 77
\bibitem[Markey \& Tayler (1974)]{martay+74} Markey, P. \& Tayler, R.J., 1974, MNRAS, 168, 505
\bibitem[Tayler (1973)]{tay+73} Tayler, R.J., 1973, MNRAS, 161, 365
\bibitem[Thompson \& Duncan (1993)]{thodun+93} Thompson, C. \& Duncan, R., 1993, ApJ, 408, 194
\bibitem[Reisenegger (2009)]{rei+09} Reisenegger, A., 2009, A\&A, 499, 557
\bibitem[Roberts (1981)]{rob+81} Roberts, P.H., 1981, Astron. Nachr., 302, 65
\bibitem[Wright (1973)]{wri+73} Wright, G.A.E., 1973, MNRAS, 161, 339
\end{thebibliography}
\end{document}